\documentclass[nofootinbib,12pt]{revtex4}
\usepackage[utf8]{inputenc}
\usepackage{graphicx}
\usepackage{amsmath}
\usepackage{amssymb}
\usepackage{color}
\usepackage{todonotes}
\usepackage{enumerate}

\begin{document}

\title{Grounding force-directed network layouts with latent space models}
\author{Felix Gaisbauer}\thanks{Corresponding author, \url{felix.gaisbauer@mis.mpg.de}}
\author{Armin Pournaki}
\author{Sven Banisch}
\author{Eckehard Olbrich}
\affiliation{Max Planck Institute for Mathematics in the Sciences \\ Inselstra\ss e 22\\ 04103 Leipzig\\Germany}\thanks{This project has received funding from the European Union’s Horizon 2020 research and innovation programme under grant agreement No 732942.}

\begin{abstract}
Force-directed layout algorithms are ubiquitously-used tools for network visualisation across a multitude of scientific disciplines. However, they lack theoretical grounding which allows to interpret their outcomes rigorously and can guide the choice of specific algorithms for certain data sets. We propose an approach building on latent space models, which assume that the probability of nodes forming a tie depends on their distance in an unobserved latent space. From such latent space models, we derive force equations for a force-directed layout algorithm. Since the forces infer positions which maximise the likelihood of the given network under the latent space model,  the force-directed layout becomes interpretable. We implement these forces for unweighted and weighted networks and spatialise different real-world networks. Comparison to existing layout algorithms (not grounded in an interpretable model) reveals that node groups are placed in similar configurations, while said algorithms show a stronger intra-cluster separation of nodes, as well as a tendency to separate clusters more strongly in retweet networks. We also explore the possibility of visualising data traditionally not seen as network data, such as survey data.
\end{abstract}

\maketitle

\section{Introduction}
\label{intro}
This contribution aims to bring together two strands of research: Latent space approaches to network analysis and force-directed layout algorithms (FDLs). FDLs are used ubiquitously for network exploration, illustration, and analysis in a wide variety of disciplines \cite{adamic2005political,mcginn2016visualizing,Conover2011,steinberg2016environmental,conover2011political,gaisbauer2021ideological,Vliet2020,Venturini2021,decuypere2020visual,pournaki2021twitter}. Nevertheless, it is still unclear how to precisely interpret node positions and corresponding patterns such as node clusters in force-directed layouts. Nor is it clear what constitutes an appropriate algorithm choice for different network data from the range of FDLs available---points also highlighted recently in \cite{jacomythesis} and \cite{Venturini2021}. We argue and show that explicit interpretability can be provided by latent space approaches, which have the goal of embedding a network in an underlying social space, and where 
link probabilities are related to proximity in this space.

To this end, we first briefly sketch how FDLs became the predominant tool for graph drawing, their connection to modularity, and their shortcomings with respect to interpretability. We also introduce latent space approaches to network analysis, and subsequently show how force terms of a new type of FDL can be derived from said latent space models, where the forces move nodes towards positions and parameters which maximise the likelihood for the network under the given model. We derive force equations for three types of networks: Unweighted networks, cumulative networks (such as the much-studied Twitter retweet networks), and weighted networks.
We present an implementation of the FDL as well as a number of real-world networks spatialised with it. We also show that existing algorithms, specifically ForceAtlas2 \cite{jacomy2014}, Fruchterman Reingold \cite{fruchterman1991graph}, and Yifan Hu \cite{hu2005efficient}, differ from the presented FDL.

\section{Latent spaces and FDLs: An intersection}

Initially, network visualisation algorithms\footnote{This paragraph largely follows \cite{jacomythesis} in its account of the history of graph drawing and (force-directed) network layout algorithms.} had been conceived to facilitate graph reading -- they were supposed to make small networks readable in the sense that paths and nodes in the network were clearly accessible, that the edges had similar lengths and that the network was drawn as symmetric as possible \cite{eades1984heuristic}. 
Progress in network science and the sudden availability of very large network data sets at the end of the millennium -- for which a comprehension of individual node positions and paths was illusory -- shifted focus: Now, networks needed to be drawn so that community structure and topological features were mediated in the layout. FDLs (partly having been developed already before this complex turn, notably in \cite{fruchterman1991graph,eades1984heuristic}) turned out to be useful and efficient tools for this task. The algorithms have in common that all nodes repel each other (the repulsive force is usually proportional to a power of of the distance between nodes, i.e. $F_r \propto d^r$), while connected nodes are additionally drawn together by their edges ($F_a \propto d^a$, $a > r$).

Noack, in a seminal work \cite{noack2009modularity}, connected FDLs to modularity, one of the most central measures of clustering in networks in use today. Roughly speaking, modularity $Q$ compares the proportion of links connecting nodes within a group of nodes with the proportion expected if the edges in the network were randomly rewired \cite{newman2004}. Community detection algorithms, such as the Louvain algorithm \cite{blondel2008}, aim to find partitions of a network that maximise this value. Noack showed that, under certain constraints, modularity can be transformed into an expression that equals the energy function of force-directed layouts. Constraints for the equivalence are \begin{enumerate}[(i)]
\item that nodes can only be placed either at the same position (then, they belong to the same cluster) or at distance 1 from each other (if not in the same cluster).
\item that FDLs operate in a space of (at least) $k-1$ dimensions, where $k$ is the number of modularity clusters (usually, FDLs embed networks in a two-dimensional space).
\item that the exponents of attractive and repulsive force should be non-negative. (Obviously, if the repulsive force has a negative exponent, placement of nodes at the same position would be impossible.)
\end{enumerate}
For FDLs, this means that if they fulfill (ii) and (iii), energy-minimal states of force-directed layouts are \textit{relaxations} of modularity maximization: They make community structure in networks visible without constraint (i) of having to sort nodes into different, fixed partitions with distance 0 or 1 from each other. They can assign continuous positions in space. Or, phrased the other way round: Modularity is then a special case of the energy function of FDLs. 

However, Noack's finding is diluted by the fact that network visualisations with FDLs are commonly restricted to two (or at most three) dimensions; and moreover, most FDLs in use today employ a negative exponent for the repulsive force. Noack also gave qualitative observations of which algorithms, even if they do not \textit{exactly} fulfill the constraints above, tend to produce results that resemble modularity clusterings. Exponents in the forces should be characterized by $a\ge 0$ and the closer to 0, the better, $r\le 0$, and $a-r\approx 1$.\footnote{ForceAtlas2 ($a = 1$, $r = -1$) is in that sense more closely related to modularity than FruchtermanReingold ($a = 2$, $r = -1$) or Yifan Hu (which uses similar forces to FruchtermanReingold, but with a multilevel algorithm).} 

The connection to modularity -- which, notably, had not been intended in the design of the algorithms -- helped give additional credibility to FDLs. All in all, this led to the widespread adoption of FDLs for network visualisation: They were not only used for illustrative purposes \cite{adamic2005political,conover2011political,Conover2011}, but also to explore and analyse network data \cite{pournaki2021twitter,Vliet2020,decuypere2020visual,mcginn2016visualizing,Venturini2021}. It is, however, unclear what information FDLs add to modularity clustering by placing the nodes in a continuous space. It has been stressed that while they have been widely used, a thorough assessment of what exactly is entailed by the produced layouts has not been provided yet \cite{jacomy2014}.\footnote{\cite{jacomythesis} departs into a somewhat different direction than the work presented here by proposing certain interventions which help interpret what is visible in FDLs that are already in use today; but both approaches try to tackle the shortcomings of FDLs that are sketched above.} And on the question of what it means that two nodes are placed close to each other by a certain FDL, answers have remained somewhat tentative: ``While in spatialized networks closer nodes \textit{tend to be} more directly or indirectly associated, no strict correlation should be assumed between the geometric distance and the mathematical distance'' \cite[p. 4]{Venturini2021} (technically, a geometric distance is of course also a mathematical distance. With the expression `mathematical distance', the authors apparently refer to a kind of graph-theoretic distance, e.g. the shortest path between two nodes). Moreover, many different types of FDLs have been developed, several of which at least approximately subsume modularity clustering.\footnote{We note here that modularity clustering is not without significant weaknesses, such as its resolution limit \cite{fortunato2007resolution} or strong degeneracies of high-scoring solutions \cite{good2010performance}.} But which one of them constitutes an appropriate choice for a certain data set at hand?\footnote{Certain quality measures to \textit{compare} network layouts have been proposed, such as the normalized atedge length \cite{noack2007unified} corresponding to the total geometric length of the edges of a network divided by the graph density and the total geometric distance between nodes. But these do not give meaning to the produced layout beyond network-immanent topological features.} 

FDLs are often implemented in easily accessible tools such as Gephi \cite{bastian2009gephi}. Not all researchers using the tool might possess the methodological training to assess the mechanics behind them. But the problems sketched above give an additional explanation for the fact that the limits or benefits of a chosen FDL are usually not discussed and ``tools such as Gephi [are often treated as] as black boxes'' \cite{bruns2013faster}.
FDLs lack \textit{grounding}, for example through an underlying model generating the forces, with which the distances between nodes could be given a rigorous interpretation, and forces could be chosen that are suitable to the network data one wants to analyse.  We propose to base a new type of FDL on latent space models of network analysis, with which network layouts can be interpreted explicitly.

Latent space approaches to (social) network analysis have been developed to infer social or political positions of actors in an underlying latent space from their interactions. They represent a class of models based on the assumption that the probability that two actors establish a relation depends on their positions in an unobserved social space \cite{Hoff2002,Handcock2007}. The social space can be constituted by a continuous space, such as an Euclidean space, or a discrete latent space, where each node is in one of several latent classes \cite{matias2014modeling}, such as the well-studied stochastic blockmodel \cite{holland1983stochastic}.  In the physics literature, models of this type were introduced under the name of spatially embedded random networks \cite{Barnett2007}. There, the Waxman model \cite{Waxman1988} and random geometric graphs \cite{Penrose2003random, Dall2002random}  were recognized as specific examples.
Recently, latent space models have been employed in the estimation of continuous one-dimensional ideological positions from social media data  \cite{Barbera2015,Barbera2015a,imai2016fast}, specifically from Twitter follower networks. The works covered large quantities of users and showed good agreement with e.g. party registration records in the United States \cite{Barbera2015}. The estimation of positions in the latent space was achieved with correspondence analysis in \cite{Barbera2015}, while in \cite{Barbera2015a}, a Bayesian method was used where the posterior density of the parameters was explored via Markov Chain Monte Carlo methods.

This is where the present work intersects: We attempt to take an alternative route in order to arrive at a specific form of force equations for FDLs. We obtain the forces on the basis of latent space models. The positions of the nodes in an assumed latent space influence the probability of ties between them -- the closer their positions, the more probable it is that they form a tie. We derive an FDL as a maximum likelihood estimator of such a model. This approach clarifies the underlying assumptions of our layout algorithm and makes the resulting layout \textit{interpretable}. We derive three different forces for three different types of networks, specifically adapted to the task of embedding them in a political space: unweighted, cumulative, and weighted networks. 
Moreover, alternative interaction models can in principle be used to develop force-directed layouts in a completely analogous way. For this, the present work can serve as a blueprint.

If one wants to take network layouts seriously, an approach highlighting the underlying assumptions of a layout and guiding its interpretation is necessary. While some might claim that the visualisation of a network only serves illustrative purposes, their wide-spread use, not only for exploration and illustration, but also visual analysis of networks \cite{conover2011political,Vliet2020,decuypere2020visual,mcginn2016visualizing,Venturini2021} underscores the necessity of this enterprise: Exploration and interpretation are, in practice, guided by force-directed layouts for many researchers from a variety of disciplines.

\section{From latent space models to force equations}
In this section, we show how force terms in a force-directed layout algorithm can be derived from latent space models of node interactions. Central to this procedure is the assumption that nodes tend to form ties to others that are close to them in a latent social space. The closer two nodes, the higher the probability that one forms a tie to the other.
Since none of the positions (as well as none of the additional parameters of the statistical model which will be introduced in the corresponding subsections) are directly observed, the statistical problem posed here is their inference. Given the underlying model, one can determine the likelihood function $L(G)$ for any observed network. The positions and parameters are then inferred via maximum likelihood estimation. In our approach, this is done by treating the negative log-likelihood as a potential energy. The minima of this potential energy are the local maximisers of the likelihood. Its derivatives with respect to the positions and parameters of the nodes can be considered as forces that move the nodes towards positions that maximise the likelihood.

We will cover three different types of directed networks: Unweighted networks, such as the follower networks covered by Barber\'a and colleagues \cite{Barbera2015,Barbera2015a}, cumulative networks (which include Twitter retweet networks), and weighted networks. Undirected networks are implicitly included as a special case where $a_{ij} = a_{ji}$ and each node only has one additional parameter $\alpha$. We will present the derivation of the forces for the unweighted case in detail. The complete derivations for the other two cases are given in Appendices \ref{ap:cumulative} and \ref{ap:weighted}.

\subsection{Unweighted networks}
Consider an unweighted graph $G=(V,E)$ with nodes $i \in V$  and edges $(i,j) \in E$. The graph can be described by an adjacency matrix $A=\{a_{ij}| a_{ij}=1 \: \text{if} \: (i,j) \in E\}$. Now let us assume that the nodes are represented by vectors $\mathbf{x}_i \in \mathbb{R}^n$ and $d_{ij}=d(\mathbf{x}_i,\mathbf{x}_j)$ denotes the Euclidean distance between $\mathbf{x}_i$ and $\mathbf{x}_j$.
For the probability of a tie between two nodes, we choose (see \cite{Hoff2002})
\begin{equation}
    p(a_{ij} = 1) = \text{logit}^{-1}(\alpha_{i} + \beta_{j} - d_{ij}^2) = \frac{\text{exp}(\alpha_{i} + \beta_{j} - d_{ij}^2)}{1+\text{exp}(\alpha_{i} + \beta_{j} - d_{ij}^2)}.
    \label{eq:model}
\end{equation}
The probability is dependent on the squared Euclidean distance between the two node positions. That the probability is dependent on the squared distance is also assumed in \cite{Barbera2015a}, while in \cite{Barbera2015,Hoff2002}, the linear distance is used. $\alpha$ and $\beta$ are additional parameters that also influence the probability of a tie. $\alpha_i$ can be interpreted as an activity parameter related to the out degree of node $i$: The higher $\alpha_i$, the higher the probability of a tie from $i$ to others. $\beta_j$ influences the probability of ties \textit{to} $j$ and influences the in degree of node $j$. The parameters allow nodes that occupy the same position in space to have different degrees -- as an example, there might be people with roughly the same political position as, say, a state leader, but it is generally unreasonable to expect that these users have the same amount of followers on social media. On the other hand, some users might simply be more active than others, hence forming more ties, while sharing a political position. 

The model introduced here has already been well-established in the literature \cite{Hoff2002,Barbera2015,Barbera2015a}. However, alternative models can be posited that might be more fitting for certain network types. In that case, the derivation lined out below can be carried out analogously.

For a given graph $G$, the likelihood function $L(G)$ can be written as the product of the probability of an edge \textit{if} there exists an edge between two nodes, and the probability of there not being an edge if not:
\begin{align}
    L(G)&=\prod_{(i,j) \in E} p(a_{ij} = 1) \prod_{(i,j) \notin E} (1- p(a_{ij} = 1)) \nonumber \\
    &= \frac{\prod_{(i,j) \in E} \exp{(\alpha_i+\beta_j-d_{ij}^2)}}{\prod_{i,j \atop i \neq j} (1+\exp{(\alpha_i+\beta_j-d_{ij}^2)})}.  
\label{eq:likelihood_unweighted}    
\end{align}
The logarithm of the likelihood is given by
\begin{align}
    LL(G):&=\log L(G)  \nonumber \\
    &=\sum_{(i,j) \in E} (\alpha_i+\beta_j-d_{ij}^2) - \sum_{i,j \atop i \neq j} \log (1+\exp(\alpha_i+\beta_j-d_{ij}^2)). 
\label{eq:loglikelihood_unweighted}    
\end{align}
If we consider the \textit{negative} log-likelihood as a potential energy, the minima of this potential are the local maximisers of the likelihood. \textit{Its (negative, once again) derivatives with respect to the positions $\mathbf{x}_i$ of the nodes can be considered as forces that move the nodes towards positions that maximise the likelihood.} For a concrete node $i'$, an attractive force is generated by node $j'$ if $i'$ establishes a tie to $j'$:
\begin{equation}
  F_{\text{att}, i'}^{j'} = \frac{\partial}{\partial \mathbf{x}_{i'}} (\alpha_{i'}+\beta_{j'}-d_{i'j'}^2)  = -2(\mathbf{x}_{i'}-\mathbf{x}_{j'}).
\end{equation}
If $j'$ also establishes a tie to $i'$, the same attractive force is applied again.
On the other hand, a rejecting force is always present for each possible tie:\footnote{Note the sign reversal in the exponent of the exponential function in the denominator in the last equivalence, which stems from $\text{exp}(\alpha_{i'}+\beta_{j'}-d_{i'j'}^2)/(1+\text{exp}(\alpha_{i'}+\beta_{j'}-d_{i'j'}^2)) = 1/(1+\text{exp}(-\alpha_{i'}-\beta_{j'}+d_{i'j'}^2))$.}

\begin{equation}
  F_{\text{rej}, i'}^{j'} = - \frac{\partial}{\partial \mathbf{x}_{i'}} \log (1+\exp(\alpha_{i'}+\beta_{j'}-d_{i'j'}^2))  = \frac{1}{1+\text{exp}(-\alpha_{i'}-\beta_{j'}+d_{i'j'}^2)} 2 (\mathbf{x}_{i'} - \mathbf{x}_{j'}).
\end{equation}
Another repulsive force on $i'$ appears for this node pair for the potential tie from $j'$ to $i'$.

The derivative of Eq. (\ref{eq:loglikelihood_unweighted}) with respect to $\alpha_{i'}$ and $\beta_{i'}$ gives us the forces on the parameters of node $i'$, such that

\begin{equation}
    F_{\alpha_{i'}}^{j'} =  a_{i'j'} - \frac{1}{1+\text{exp}(-\alpha_{i'} - \beta_{j'} + d_{i'j'}^2)} = a_{i'j'} - p(a_{i'j'} = 1)
\end{equation}
and 
\begin{equation}
    F_{\beta_{i'}}^{j'} =  a_{j'i'} - \frac{1}{1+\text{exp}(-\alpha_{j'} - \beta_{i'} + d_{i'j'}^2)} = a_{j'i'} - p(a_{j'i'} = 1).
\end{equation}

The sum over all forces on the single parameters yields the difference between the actual and the expected in/out degree. In the equilibrium state, where this sum yields 0, the observed in/out degree of nodes equals the one expected under the model of Eq. (\ref{eq:model}):
\begin{equation}
    \sum_{j'} F_{\alpha_{i'}}^{j'} = d_i^{\text{out}} - \langle d_i^{\text{out}} \rangle,
\end{equation}

\begin{equation}
    \sum_{j'} F_{\beta_{i'}}^{j'} = d_i^{\text{in}} - \langle d_i^{\text{in}} \rangle.
\end{equation}

\subsection{Cumulative networks}

Force equations can also be derived for networks which are constituted by a number of binary signals between nodes -- for example, when users of an online platform create several posts, each of which can be taken up by others (e.g. through liking or sharing the respective post). A much-studied case are Twitter retweet networks, which are frequently employed to investigate opinion groups on the platform \cite{Conover2011,conover2011political}. 

We consider, for each action $k$ initiated by a node $j$ (e.g. a tweet), an unweighted graph $G_{j}^k=(V,E_{j}^k)$ with nodes $i \in V$  and edges $(i,j) \in E_{j}^k$, where an edge means that $i$ has formed a tie to $j$ upon action $k$. The graph for each $k$ can be described by an adjacency matrix $A_{j}^k=\{a_{ij}| a_{ij}=1 \: \text{if} \: (i,j) \in E_{j}^k\}$. This constitutes an $m$-star graph with $m=|E_j^k|$.

Analogously to the unweighted case, we assume the probability of establishing \textit{a single tie upon action $k$ from user $i$ to user $j$} with
\begin{equation}
    p(a_{ij}^k = 1) = \frac{1}{1+\text{exp}(-\alpha_i - \beta_{jk}+d_{ij}^2)},
\end{equation}
where where each action $k$ of $j$ has its own parameter $\beta_{jk}$ which affects the in degree of $j$.

The log-likelihood for the cumulative 
network can be written as
\begin{equation}
    LL(G) = \sum_{j} \sum_{k=1}^{m_j} \Big( \sum_{(i,j)\in E_{jk}}(\alpha_i + \beta_{jk} - d_{ij}^2) - \sum_{i \atop i\neq j} \text{log}(1+e^{-\alpha_i -\beta_{jk}+d_{ij}^2}) \Big)
\end{equation}
Here, in addition to user pairs, we sum over all actions $k$. The derivation of forces for this case is largely analogous to the unweighted case and can be found in  Appendix \ref{ap:cumulative}, along with the concrete force equations.

\subsection{Weighted networks}
\label{subsec:weighted}

So far, we have assumed a binary signal between node pairs -- e.g. whether an individual follows another one or not, or whether someone shares certain content of another individual or not. Extending the model to the non-binary case can be achieved by exchanging the ordinary logit model (Eq. (\ref{eq:model})) with an ordered logit or proportional adds model.

There, a response variable has levels $0, 1,..., n$ (e.g.: people rate their relationships to others on a scale from 0 to 6, or similar).
We consider the general case of weighted networks with finite weights that can be transformed into natural numbers (with 0), i.e. an adjacency matrix $A=\{a_{ij}=k | k \in \mathbb{N}_0\}$. The probability of the variable being greater than or equal to a certain level $k$ is given by \cite{harrell2015regression,kleinbaum2002logistic}: 

\begin{equation}
     p(a_{ij} \geq k) = \frac{1}{1 + \text{exp}(-c_k - \alpha_i-\beta_j+d_{ij}^2)},
\end{equation}
where $k = 0,1,...,n.$ ($c_0 = \infty$, $c_{n+1} = -\infty$.) The probability of $a_{ij}$ equal to a certain $k$ is given by
\begin{equation}
    p(a_{ij} = k) = P(a_{ij} \geq k) - P(a_{ij} \geq k+1).
\end{equation}

The likelihood $L(G)$ is given by
\begin{equation}
    L(G) = \prod_{i,j} (\frac{1}{1 + \text{exp}(-c_{a_{ij}} - \alpha_i-\beta_j+d_{ij}^2)} - \frac{1}{1 + \text{exp}(-c_{a_{ij}+1} - \alpha_i-\beta_j+d_{ij}^2)}),
\end{equation}

and the log-likelihood by
\begin{equation}
    LL(G) = \sum_{i,j\atop i\neq j} \text{log}(\frac{1}{1 + \text{exp}(-c_{a_{ij}}  - \alpha_i-\beta_j+d_{ij}^2)} - \frac{1}{1 + \text{exp}(-c_{a_{ij}+1}  - \alpha_i-\beta_j+d_{ij}^2)}). \label{eq:logl-weighted}
\end{equation}

The force equations derived from the log-likelihood (for the case of a three-point scale) can be found in Appendix \ref{ap:weighted}. 
Potential applications of a visualisation with these forces are manifold. In smaller data sets, a non-binary signal between nodes, e.g. rating of the relationships between individuals of a social group, might be given between all node pairs. But often, there might be cases for which a subset of individuals (say, politicians, public figures, etc.) or items (e.g. the importance of political goals, technologies, etc.) are rated by others. Then, only the rating individuals receive an $\alpha$, while the rated just have a $\beta$-parameter.
The interpretation of the parameters $\alpha$ and $\beta$ might need adjustment: They now rather refer to the tendency of individuals to give/receive rather high/low ratings.

\section{Implementation and validation}

The force-directed layout algorithm was implemented in JavaScript, building upon the d3-force library \cite{d3-force}. There, force equations are simulated using a velocity Verlet integrator \cite{Verlet1967,Swope1982}. A ready-to-use implementation, which we call \textit{Leipzig Layout}, is available under \url{https://github.com/pournaki/leipzig-layout}.\footnote{For the moment, this implementation is restricted to unweighted graphs. An extension for weighted and cumulative graphs will be published on the same repository.} It builds upon the force-graph library \cite{vasturiano_forcegraph} to interactively display the graph and the evolution of node positions in the simulation of forces. Note that, at the current state, this layout tool works reasonably fast for networks below 10,000 links.

\begin{figure}
    \centering
    \includegraphics[width = 1.03\textwidth]{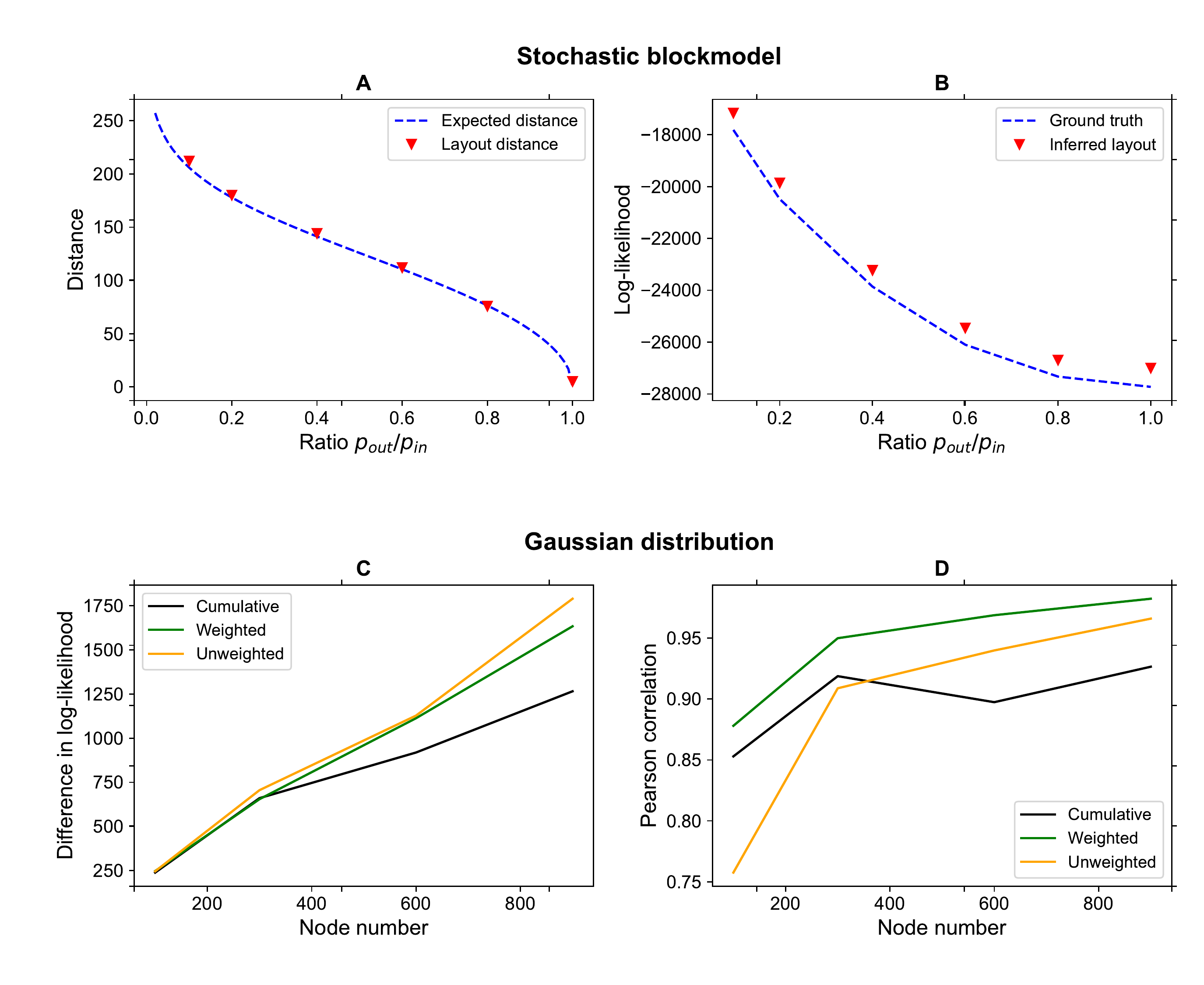}
    \caption[Validation]{Expected distance of an SBM (two blocks, 100 nodes each) with varying $p_{\text{out}}$ compared to the distance between the center of mass of the clusters in the proposed layout algorithm, averaged over 5 runs (A). Not only is the inferred distance by the force-directed layout algorithm nearly identical to the expected one throughout, but the log-likelihood of the inferred latent space surpasses the ground truth in all cases (B). The difference between inferred log-likelihood and the log-likelihood of the ground truth for a Gaussian distribution of two groups of nodes ($\sigma = 1/12$, $d = 5/6$, averaged over 3 runs) is given in C. In all cases, the log-likelihood of the inferred latent space surpasses the ground truth (i.e. them difference in log-likelihood is positive). Still, similarity between ground truth and inferred distances between nodes is high (increasing with number of nodes), as is visible in the average Pearson correlation between distance matrices (D).}
    \label{fig:validationsbm}
\end{figure}

Validation for the unweighted case was performed by testing the agreement with the expected distance of a stochastic block model (SBM) of two blocks with varying $p_{\text{out}}$ and $p_{\text{in}} = 0.5$. In the above model, the expected distance can be computed by placing the nodes of each block on the same point in space, and then choosing the distance $d$ between the two blocks so that $p_{\text{out}} = \frac{1}{1+\text{exp}(-d^2)}$ ($\alpha$ and $\beta$ are set to 0 so that the expected degree for all nodes in one block is the same).
With this underlying latent space, one can draw a network according to the given probabilities and let the layout algorithm infer the latent space again. Averaged over 5 runs and for 100 nodes per block, we observe that the inferred distance of the centers of mass of the blocks are nearly identical to the expected one (Fig. \ref{fig:validationsbm} A), and the log-likelihood of the inferreds latent space surpasses the one of the actually drawn one in all cases (Fig. \ref{fig:validationsbm} B). 

Moreover, we compare the log-likelihood of the inferred latent space with the one of the actually drawn network from a Gaussian distribution of two groups of nodes with a $\sigma$ of 1/12 and a distance of 5/6 between the groups with varying node number, averaged over three runs. In all cases, the log-likelihood of the inferred latent space is higher than the ground truth (Fig. \ref{fig:validationsbm} C). Still, similarity to the ground truth distances between nodes was high throughout, which we assessed with a Mantel test \cite{mantel1967detection}. Pearson correlation between distance matrices can be inspected in Fig. \ref{fig:validationsbm} D, the average z-score is reported in Appendix \ref{ap:zscore}.


\section{Real-world networks \label{sec:real-w}}
\begin{figure}
   \noindent
  \makebox[\textwidth]{
    \includegraphics[width = \textwidth]{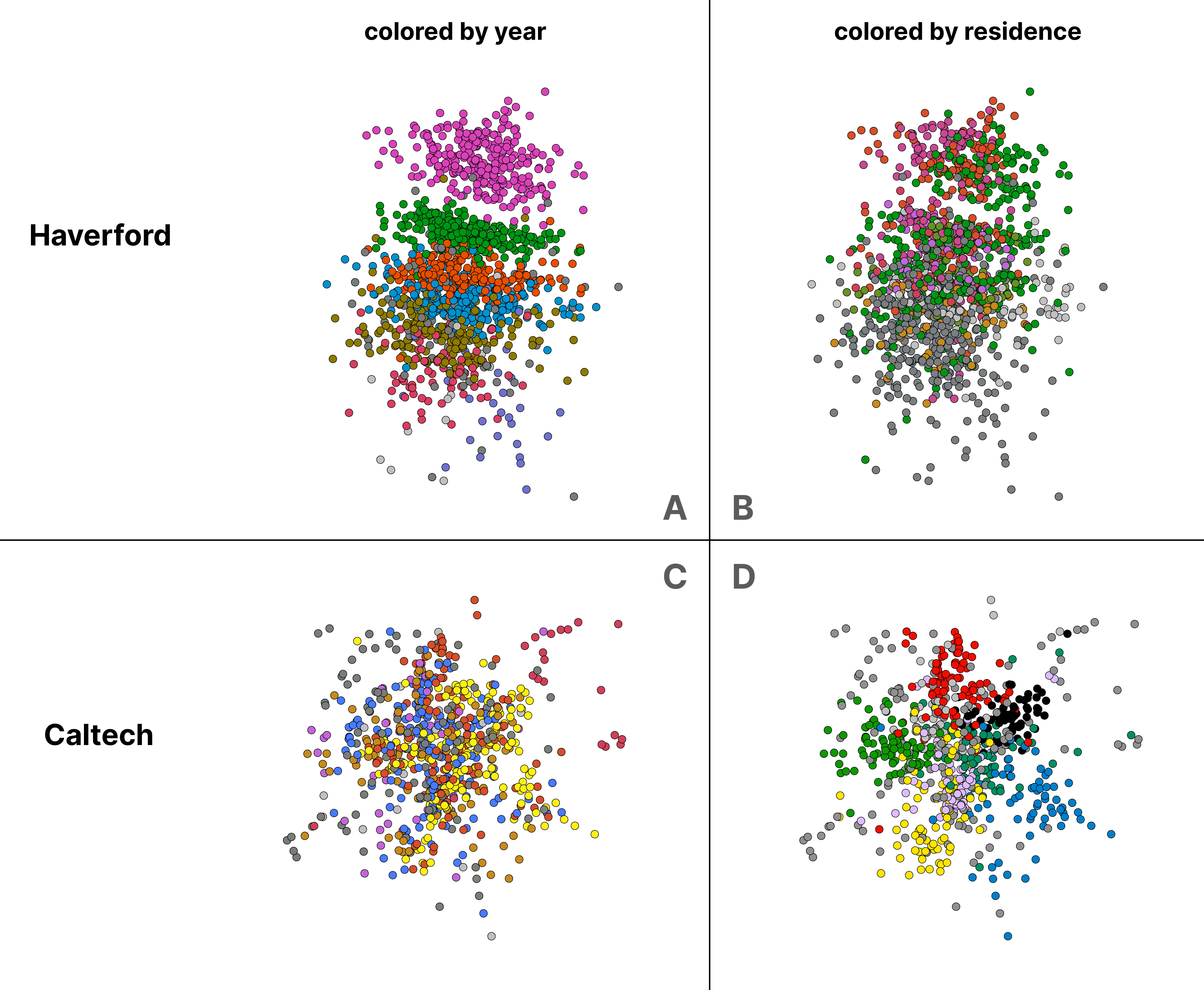}}
    \caption[Haverford friendship network]{Friendship network of students of Haverford University (top) and California Institute of Technology (bottom), colored by year (left) and residence (right). The spatialisation of the former layers students by year (A, chronologically ordered from top to bottom, with first-year students colored pink, second-year students colored green, etc.; dark grey nodes correspond to students whose year is unknown). First-year students are visually separated from the others, while the layout becomes denser if students have been at university for a longer time. In B, it is also visible that first-year students show a higher tendency to mix with others they share residency with (dark grey: dorm unknown). For Caltech, the network out of the Facebook100 data set with the highest assortativity with respect to residence, nodes are visibly placed according to dorm membership (D, dark grey: year/dorm unknown), and less so with respect to year (C).}
    \label{fig:facebook}
\end{figure}

Next, we use Leipzig Layout to spatialise several real-world networks: Undirected Facebook friendship networks, the directed Twitter follower network of the German parliament, the retweet network of Twitter debate surrounding the publication of a letter on free speech by Harper's magazine, and a survey on different types of energy-generating technologies.
\paragraph{Facebook100: Haverford \& Caltech}
The Facebook100 data set consists of online social networks collected from the Facebook social media platform when the platform was only open to 100 universities in the US \cite{traud2012social}. The data set contains social networks of students of particular universities with quite rich metadata (e.g. gender, year, residence, or major). We analyse friendship networks -- undirected networks where a tie between users represents that both have agreed to connect with each other as `friends' on the platform. 
We spatialise the friendship network of Haverford University in Fig. \ref{fig:facebook} A and B (links have been omitted for better accessibility). On the left, it is visible that students are spatially layered according to their year by the layout algorithm. The first-year students are visually separated from the others. The layout becomes denser for students who have been at the university for a longer time. The layers are ordered chronologically. It seems that if students form cross-year ties, they tend to connect to others from adjacent cohorts.
The local assortativity distribution with respect to residence of the students of Haverford has been analysed in detail in \cite{peel2018}. There, it was found that first-year students tend to form ties to other students from their dorms, while students from higher years show less of a tendency to mix only with others they share residence with. This behavioral pattern can also be discerned in the spatialisation (Fig. \ref{fig:facebook} B): For the first-year students, the students sharing a dorm tend to be placed rather close to each other, while for students from higher years, this is not the case. 
As a complement, we spatialise the network for the university with the highest overall assortativity with respect to dormitory in the data set: Caltech. There, the students' year does not influence Facebook friendship to a large extent; rather, students' friendships are more strongly guided by their residence \cite{red2011comparing}. This is reproduced by Leipzig Layout: Fig. \ref{fig:facebook} D shows that students that share residence are visibly placed close to each other. On the other hand, students are less strongly grouped according to their university year (C).

\paragraph{German parliament: Twitter follower network}

\begin{figure}
   \noindent
  \makebox[\textwidth]{
    \includegraphics[width = 1.1\textwidth]{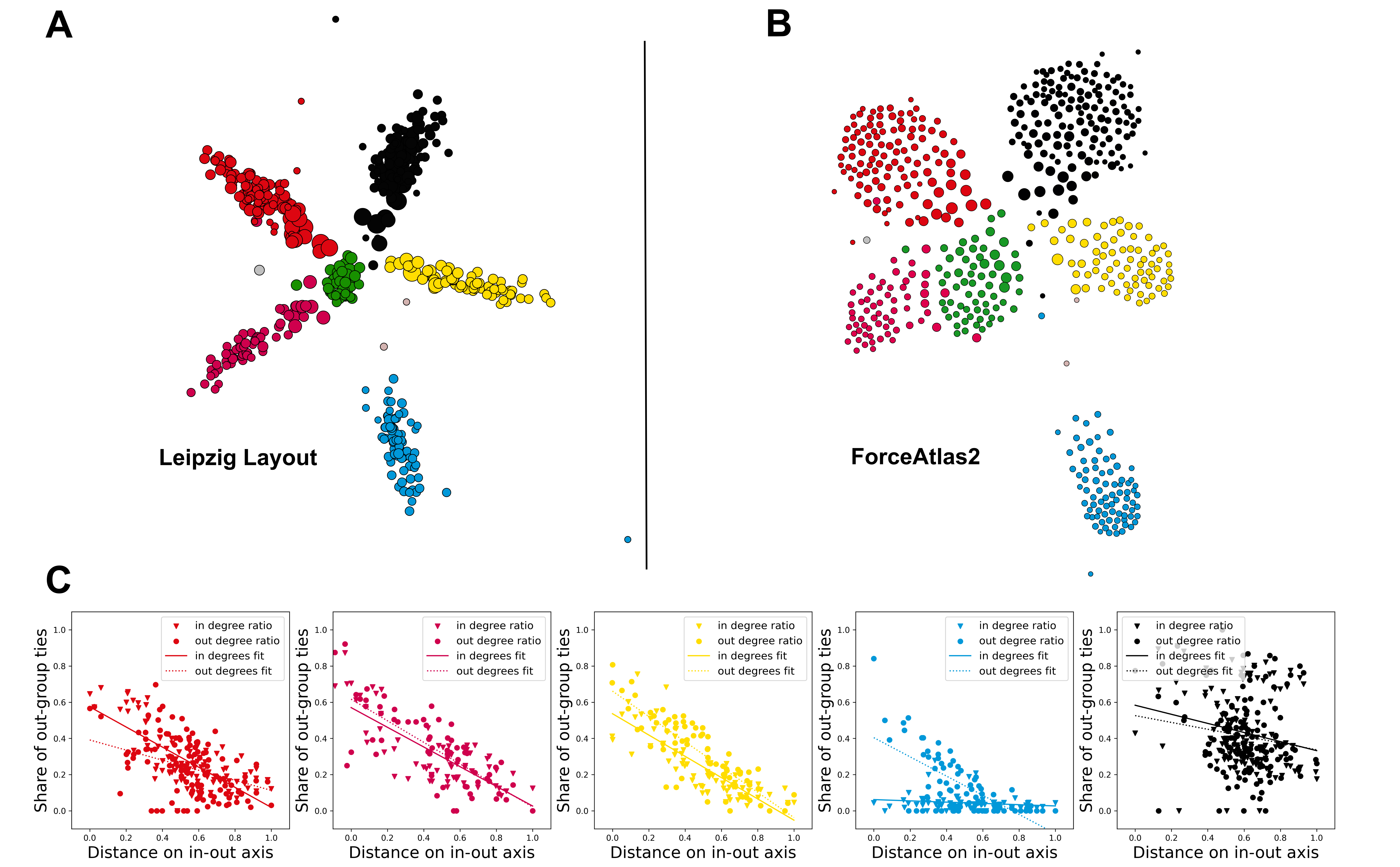}}
   \caption[German parliament follower network]{Leipzig Layout of the follower network of all German deputies that have a Twitter account (A). Members are colored according to their party and node size corresponds to overall node degree. Clear division between parties, as well as a stronger division between the right-wing party \textit{AfD} and the other parties is visible. All parties except the \textit{Greens} are arranged on a one-dimensional axis. This is explained by a difference in cross-party ties between politicians of the same party: The further out a member on the party-internal axis, the fewer cross-party ties to and from them have been established (except for the \textit{AfD}, which does not receive many ties from other parties no matter where the users are placed) (C, colored according to parties, linear fits included). ForceAtlas2, in comparison, has a stronger separation of nodes within party clusters due to its rejecting force being proportional to $d^{-1}$ (B).}
    \label{fig:bt-follow}
\end{figure}

While \cite{Barbera2015a,Barbera2015} aim for the estimation of one-dimensional ideological positions of politicians (and their followers), the FDL proposed here embeds nodes in a two-dimensional space. We spatialise the Twitter follower network of all members of the German parliament that have an active Twitter account in Fig. \ref{fig:bt-follow}. The parties (members colored according to their typical party color) are quite visibly separated. They are located along a circle that quite accurately mirrors the political constellation in federal German politics. The center-left to center-right parties (\textit{SPD}, \textit{Bündnis 90/Die Grünen} (Green party), \textit{CDU/CSU}) are positioned between \textit{Die Linke} (Left party) and the market-liberal \textit{FDP}. The \textit{AfD} (blue), a right-wing populist party with which collaboration has been ruled out by all other parties, accordingly occupies a secluded area. Interestingly, within parties, a one-dimensional arrangement is visible (except for the \textit{Greens}). This mirrors the amount of cross-party ties, as well as the users' activity on Twitter: The further out on an axis between the innermost and outermost party member users are placed (outliers excluded), the smaller their share of ties to other parties (Fig. \ref{fig:bt-follow} C). 
The central and densely packed placement of the Greens can be explained by the fact that they tend to use Twitter quite homogeneously (see Appendix \ref{ap:networks}) in the sense that there are no users which are inactive or lack followers: Each of them (except one) has an in/out degree of at least 50. Moreover, the party members are followed and follow all other parties (except for the \textit{AfD}) in a well-balanced fashion. $\alpha$ and $\beta$ are generally correlated with out and in degree of the nodes (again, see Appendix \ref{ap:networks}). They exhibit a very pronounced linear correlation for the Greens.

This layout also illustrates the difference between ForceAtlas2 (see (B) in Fig. \ref{fig:bt-follow}) and the layout algorithm at hand here: ForceAtlas2 incorporates a rejecting force between node pairs proportional to $d_{ij}^{-1}$, which leads to a stronger separation of nodes within clusters. Hence, while the overall arrangement of parties is similar to Leipzig Layout, parties themselves are more strongly spaced out. A comparison to spatialisations with the algorithms Yifan Hu and FruchtermanReingold can also be found in Appendix \ref{ap:networks}. 

Moreover, we observe that several minima are inferred by both Leipzig Layout and the other FDLs -- depending on the initial positions of the nodes. While the existence of different local minima is a general problem of FDLs, one can simply select the outcome with the highest likelihood with the present approach -- a further advantage of an FDL grounded in an underlying model. A different, but less likely local minimum inferred with Leipzig Layout is also presented in Appendix \ref{ap:networks}. The more likely minimum, which is displayed in Fig. \ref{fig:bt-follow}, is also the politically more plausible one: In Appendix \ref{ap:networks}, \textit{SPD} is placed closer to \textit{FDP} than \textit{CDU/CSU}, while the latter two parties have more commonalities (especially when it comes to economic policy).

\begin{figure}

    \includegraphics[width = \textwidth]{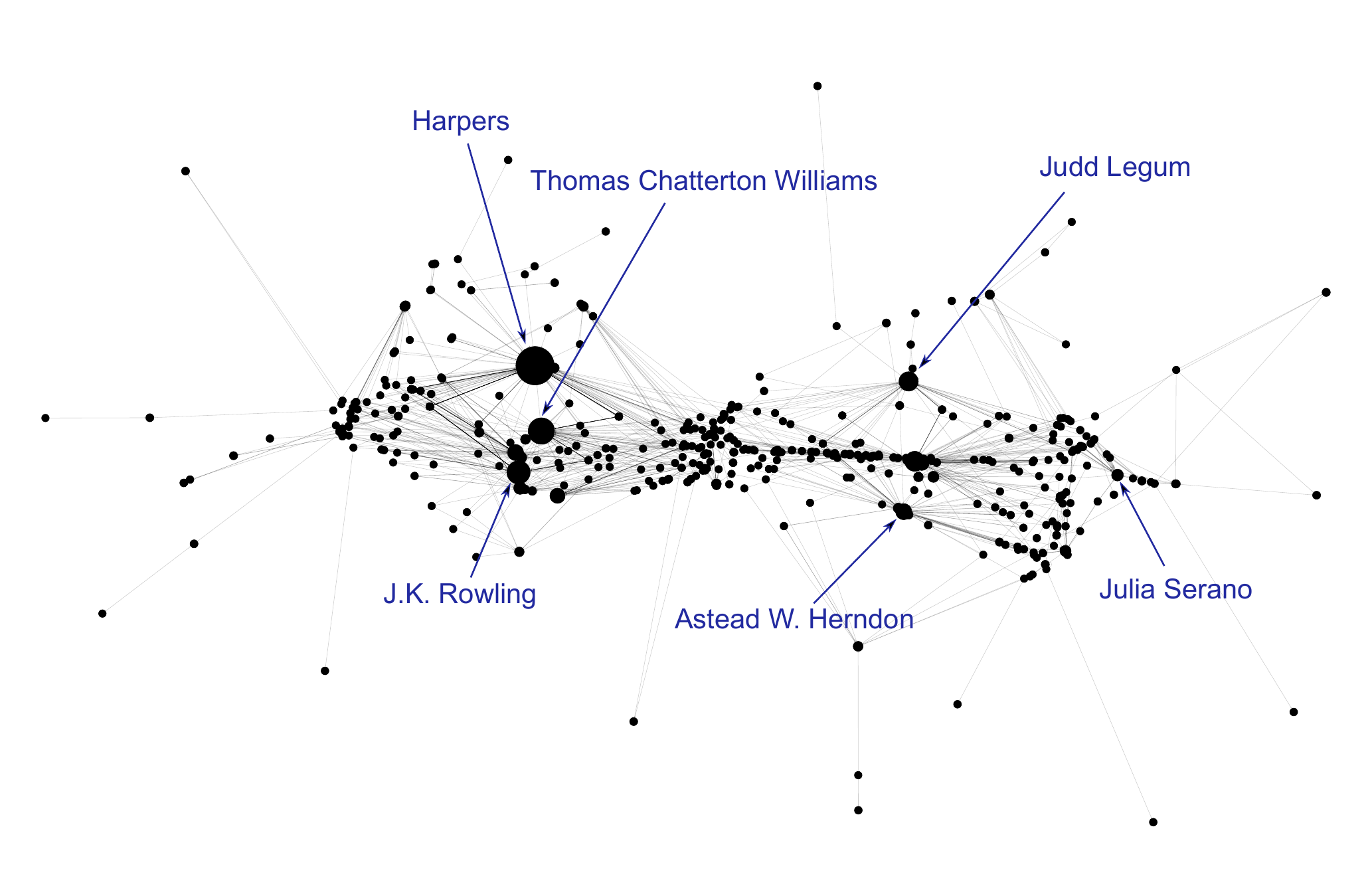}
    \caption[Retweet network for Harper's letter]{Retweet network of Twitter debate about a letter on free speech published by Harper's magazine (node size proportional to in degree). A two-camp division is visible, where the left pole includes the magazine as well as prominent signees, while the right pole contains critics.}
    \label{fig:harpers}
\end{figure}

\paragraph{Retweet network: Harper's letter}
In July 2020, Harper's magazine published an open letter signed by 153 public figures defending free speech which they saw endangered by `forces of illiberalism.' Not only Donald Trump was denounced as contributing to illiberalism, but also some groups who advancing ``racial and political justice,'' who had ``intensified a new set of moral attitudes and political commitments that tend to weaken our norms of open debate and toleration of differences in favor of ideological conformity'' \cite{harpers}. On Twitter, the letter was controversially discussed subsequently (see also \url{https://blog.twitterexplorer.org/post/harpers_letter/}).
The layout of the retweet network reproduces a division between critics and supporters of the letter: On the left side of Fig. \ref{fig:harpers}, the account of Harper's magazine as well as prominent signees such as Thomas Chatterton Williams and Joanne K. Rowling are visible, while the right pole includes critics of the letter and its signees, such as Judd Legum, Astead W. Herndon and Julia Serano. Serano, a transgender activist, criticized that what the signees referred to as `free speech' has prevented marginalized groups from speaking out, and accused Rowling of having spread disinformation about trans children. That she was voicing rather specific criticism which aimed towards certain signatories of the letter is mirrored in her position close to the margin of the inferred space. Legum and Herndon are placed closer to the center: Legum noted in a relatively nuanced critique that the signees of the letter are not silenced in any way, while Herndon published several ironical tweets about the letter. Interestingly, the division of clusters visible in the layout is not as pronounced as in the spatialisation of the network with ForceAtlas2 and Yifan Hu (see Fig. \ref{fig:harpers-comp} in Appendix \ref{ap:networks}), a finding that calls for further systematic investigation.

\paragraph{Survey data}

With the weighted layout, not only generic network data can be spatialised, but also surveys: There, evaluated items as well as respondents are nodes, and forces only exist between items and individuals.

\begin{figure}
    
    \includegraphics[width = \textwidth]{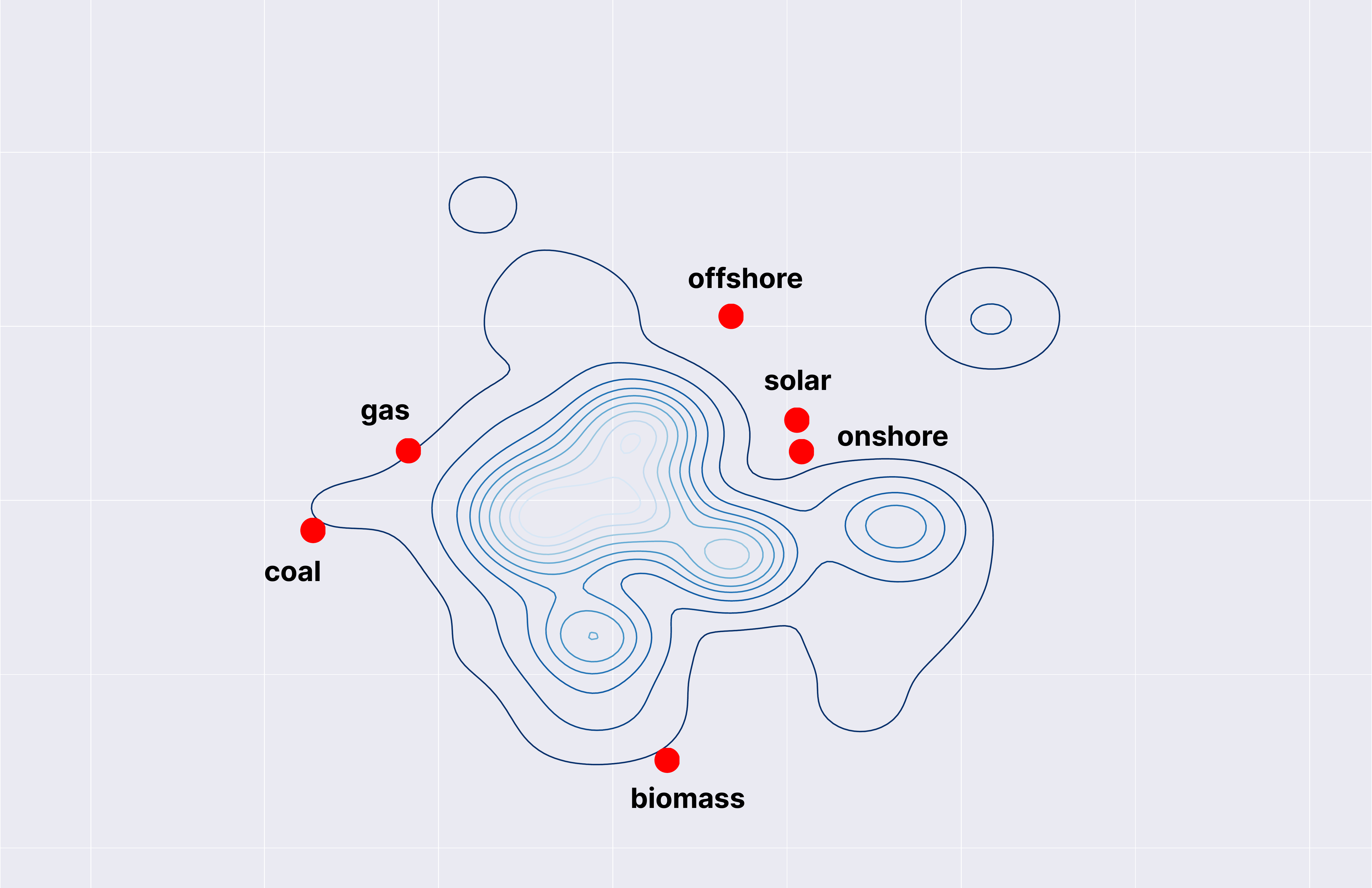}
    \caption[Survey on energy-generating technologies]{visualisation of a survey on six different energy-generating technologies. The distribution of respondents is plotted as a density in the background (the lighter, the denser they are distributed in an area). Respondents are distributed close to gas, renewable energy-generating technologies, and between them. Two technological axes are visible: One from coal and gas to the renewables, and one among technologies using renewable sources of energy, with onshore and solar occupying central positions, while offshore and biomass are located opposite of each other.}
    \label{fig:energy}
    
\end{figure}

In Fig. \ref{fig:energy}, we visualize a survey where respondents were asked about their attitude towards six different energy-generating technologies \cite{shamon2019changing}. The responses represent the initial attitudes of respondents with respect to the technologies before being confronted with several pro and counter arguments. Responses were initially given on a nine-point scale, which was aggregated to a three-point scale for visualisation. Gas and coal power stations, onshore and offshore wind stations, biomass power stations, and open-space photovoltaics (which we refer to as solar in Fig. \ref{fig:energy}).

The distribution of respondents over the inferred space, given by a density plot (the lighter the color, the more respondents lie in a region of the layout), shows that the vast majority of respondents is located between gas and onshore, solar and offshore energy-generating technologies, while coal is placed far away from most respondents. 

Several density peaks exist which correspond to respondents with similar response profiles: One between gas and biomass, one placed rather centrally between gas, offshore, solar, onshore and biomass, two between biomass and solar/onshore, and one at the margin of the space, but closest to offshore and solar/onshore technologies. 
Even more interesting is the arrangement of technologies themselves, since it shows that collectively, response profiles of individuals create two orthogonal axes along which technologies are placed: One axis is visible from renewables towards technologies relying on fossil sources of energy (gas and coal). On the other hand, renewable sources of energy are distributed along an perpendicular axis. Onshore and solar occupy central positions there, while offshore and biomass are located opposite of each other. 
The respondents' distribution and the arrangement of technolgies are in line with the average ratings and rating correlations between individuals (see Appendix \ref{ap:correlations}). Average ratings for coal are reported to be significantly lower than for any other technology in \cite{shamon2019changing}, gas receives a neutral rating, renewable technologies (offshore, onshore, solar, biomass) are rated positively on average. Biomass receives the lowest average rating of the renewables, which is reflected in the distribution of respondents. Biomass has, among the renewables, the weakest correlations with the other renewables. On the other hand, ratings are not negatively correlated with coal or gas. This is mirrored by its placement in Fig. \ref{fig:energy}.

\section{Discussion}

While FDLs are frequently employed for network visualisation across a variety of scientific disciplines, they lack theoretical grounding which allows to interpret their outcomes rigorously. We have presented a path towards interpretable FDLs based on latent space models. We have derived force equations for Leipzig Layout, a FDL that serves as a maximum likelihood estimator of said models. We have posited three variants of the FDL, which are applicable to unweighted, cumulative, and weighted networks, respectively. Exemplary spatialisations of several real-world networks show that important properties of the networks (assessed through different network measures) are reflected by node placement. Moreover, commonalities with, but also differences to existing FDLs have been pointed out: The latter tend to exhibit a stronger separation of nodes within tightly connected clusters, and, for the cumulative case, also between each other. 

The new type of FDL presented here makes the assumptions it is based on -- the underlying latent space model -- explicit, and hence constitutes an attempt to put FDLs on a more rigorous scientific basis. Latent space models are well-established in the estimation of ideological positions on the basis of (social) networks. In most cases, the ideology estimates have been carried out on a one-dimensional axis. Leipzig Layout infers a two-dimensional latent space.

The model chosen here can, if found necessary for certain network types, be replaced by alternative interaction models. For this purpose, the derivation of forces above can serve as a blueprint. The present approach might also be used to motivate parameter choices for existing FDLs, such as ForceAtlas2: The degree of influence of edge weights there, for example, can be arbitrarily chosen. But the choice could be guided by agreement with the weighted case of the algorithm implemented here.

The spatialisation of survey data presented above points beyond traditional usage of FDLs, an avenue which should be explored further. We note here that recent work in machine learning also used force-directed layouts that were derived as gradients of an objective function: Stochastic neighbor embedding (SNE) \cite{hinton2002stochastic} and in the sequel t-SNE \cite{vanderMaaten2008visualizing} and UMAP \cite{mcinnes2020umap} used this technique to embed a graph in a lower dimensional space. In these cases, the objective function was not the likelihood of a statistical model but the KL-divergence between two probability distributions.

Limitations remain: The convergence of FDLs to local minima is a problem that cannot be solved by the present approach. The follower network presented in Fig. \ref{fig:bt-follow}, for instance, possesses several equilibria for which the parties were allocated in different order, both for Leipzig Layout as well as the three algorithms it was compared with. Nevertheless, the underlying model of Leipzig Layout allows a comparison of the log-likelihood of several equilibria, out of which the most likely can then be chosen.
Moreover, the role of dimensionality for the outcomes of latent space inference in general has not been studied systematically \cite{matias2014modeling}. For network visualisation, an extension to a three-dimensional latent space would be of interest in comparison to the two-dimensional case studied here.


\bibliography{retweet}

\appendix
\section{Force derivation: Cumulative networks \label{ap:cumulative}}
For cumulative networks, we stipulate the probability of establishing \textit{a single tie upon action $k$ from user $i$ to user $j$} by
\begin{equation}
    p(a_{ij}^k = 1) = \frac{1}{1+\text{exp}(-\alpha_i - \beta_{jk}+d_{ij}^2)},
\end{equation}
The derivation of the forces for this case is analogous to the unweighted case, except that we sum over tweets instead of user pairs.

The log-likelihood for a given network can be written as
\begin{equation}
    LL(G) = \sum_{j} \sum_{k=1}^{m_j} \Big( \sum_{(i,j)\in E_{jk}}(\alpha_i + \beta_{jk} - d_{ij}^2) - \sum_{i \atop i\neq j} \text{log}(1+e^{-\alpha_i -\beta_{jk}+d_{ij}^2}) \Big)
\end{equation}

The force on the position of node $i'$ exerted by node $j'$ is given by
$$
    \partial_{x_{i'}}^{j'} LL(G) = 2(\mathbf{x}_{i'}-\mathbf{x}_j) \bigg[ \sum_{k=1}^{m_{j'}} \Big( -a_{i'j'} + \frac{1}{1+\text{exp}(-\alpha_{i'}-\beta_{j'k}+d_{i'j'}^2)} \Big) + 
$$
\begin{equation}
    \sum_{k=1}^{m_{i'}} \Big( -a_{j'i'} + \frac{1}{1+\text{exp}(-\alpha_{j'}-\beta_{i'k}+d_{i'j'}^2)} \Big) \bigg].
\end{equation}

The force on $\alpha_{i'}$ exerted by $j'$ is given by
\begin{equation}
    \partial_{\alpha_{i'}}^{j'} LL(G) = \sum_{k=1}^{m_{j'}} \Big( a_{i'j'} - \frac{1}{1+\text{exp}(-\alpha_{i'}-\beta_{j'k}+d_{i'j'}^2)} \Big),
\end{equation}
the force on each $\beta_{i'k}$ due to $j'$ by 
\begin{equation}
    \partial_{\beta_{i'k}}^{j'} LL(G) = a_{j'i'} - \frac{1}{1+\text{exp}(-\alpha_{j'}-\beta_{i'k}+d_{i'j'}^2)}.
\end{equation}

\section{Force derivation: Weighted networks \label{ap:weighted}}
A possibility of extending the model to weighted graphs, i.e. an adjacency matrix $A=\{a_{ij}=k | k \in \mathbb{N}_0\}$, is the ordered logit or proportional odds model. There, the response variable has levels $0,1,..., n$ (e.g.: people rate their relationships to others on a scale from 0 to 6). The probability of the variable being greater than or equal to a certain level $k$ is given by: 
\begin{equation}
    p(a_{ij} \geq k) = \frac{1}{1 + \text{exp}(-c_k - \alpha_i-\beta_j+d_{ij}^2)}
\end{equation}
($c_0 = \infty$, $c_{n+1} = -\infty$). The probability of $a_{ij}$ equal to a certain $k$ is given by
\begin{equation}
    p(a_{ij} = k) = p(a_{ij} \geq k) - p(a_{ij} \geq k+1).
\end{equation}

The log-likelihood of the graph is then given by
\begin{equation}
    LL(G) = \sum_{i,j\atop i\neq j} \text{log}(\frac{1}{1 + \text{exp}(-c_{a_{ij}}  - \alpha_i-\beta_j+d_{ij}^2)} - \frac{1}{1 + \text{exp}(-c_{a_{ij}+1}  - \alpha_i-\beta_j+d_{ij}^2)}) \label{eq:logl}
\end{equation}

As the simplest example, we turn to the case with three levels, where the probabilities are given by\footnote{In practice, one might often encounter data sets where individuals have rated a limited set of others, e.g. politicians, which themselves might not participate in the rating. This yields a bi-partite graph for which $\beta$ only applies to the rat\textit{ed} individuals, while $\alpha$-values are present only for the rat\textit{ing} individuals.}

\begin{equation}
    p(a_{ij} = 0) = 1-\frac{1}{1+\text{exp}(-c_1-\alpha_i-\beta_j+d_{ij}^2)}
\end{equation}
\begin{equation}
    p(a_{ij} = 1) = \frac{1}{1+\text{exp}(-c_1-\alpha_i-\beta_j+d_{ij}^2)}-\frac{1}{1+\text{exp}(-c_2-\alpha_i-\beta_j+d_{ij}^2)}
\end{equation}
\begin{equation}
    p(a_{ij} = 2) = \frac{1}{1+\text{exp}(-c_2-\alpha_i-\beta_j+d_{ij}^2)}
\end{equation}

The derivative of Eq. (\ref{eq:logl}) with respect to $\mathbf{x}_i$, $\alpha_i$ and $\beta_i$ gives us the forces on the position and the parameters of node $i$. Additionally, the derivative with respect to $c_1$ and $c_2$ estimates the cut points. We introduce the following abbreviations:
$$ C_1 = -c_1-\alpha_{i'}-\beta_{j'}+d_{i'j'}^2,
$$
and 
$$ C_2 = -c_2-\alpha_{i'}-\beta_{j'}+d_{i'j'}^2.
$$
The first part of the force on the position of node $i'$ exerted by node $j'$ is given by
\begin{equation}
    \partial_{x_{i'}} \text{log(p}(a_{i'j'} = 0)) = 2(\mathbf{x}_{i'}-\mathbf{x}_{j'})\frac{e^{C_1}}{(1+e^{C_1})^2-(1+e^{C_1})} = 2(\mathbf{x}_{i'}-\mathbf{x}_{j'})\frac{1}{1+e^{C_1}}
\end{equation}
if $a_{i'j'} = 0$. If $a_{i'j'} = 1$, the force is given by
\begin{equation}
    \partial_{x_{i'}} \text{log(p}(a_{i'j'} = 1)) = 2(\mathbf{x}_{i'}-\mathbf{x}_{j'})
    \frac{(1+e^{C_2})^{-2}e^{C_2} - (1+e^{C_1})^{-2}e^{C_1}}{(1+e^{C_1})^{-1} - (1+e^{C_2})^{-1}}.
\end{equation}
Or, if $a_{i'j'} = 2$,
\begin{equation}
    \partial_{x_{i'}} \text{log(p}(a_{i'j'} = 2)) = -2(\mathbf{x}_{i'}-\mathbf{x}_{j'}) \frac{1}{1+e^{-C_{2}}}.
\end{equation}

For the second part of the force on $i'$, caused by $a_{j'i'}$, one simply needs to take the appropriate term out of the three above and switch $i'$ and $j'$ for $\alpha$ and $\beta$.

The force on $\alpha$ caused by $j'$ is given by 
\begin{equation}
    \partial_{\alpha_{i'}} \text{log(p}(a_{i'j'}=0)) =- \frac{1}{1+e^{C_1}}
\end{equation}
or
\begin{equation}
    \partial_{\alpha_{i'}} \text{log(p}(a_{i'j'}=1)) = 
    \frac{(1+e^{C_1})^{-2}e^{C_1} - (1+e^{C_2})^{-2}e^{C_2}}{(1+e^{C_1})^{-1} - (1+e^{C_2})^{-1}}
\end{equation}
or
\begin{equation}
        \partial_{\alpha_{i'}} \text{log(p}(a_{i'j'}=2)) = \frac{1}{1+e^{-C_2}}.
\end{equation}
Now, there is no second force part -- this is the only contribution by the pair $i'$ and $j'$. The force on $\beta_{i'}$ is given by the analogous term where $i'$ and $j'$ are, again, switched for $\alpha$ and $\beta$, and the level of $a_{j'i'}$ is considered.

The forces on $c_1$ and $c_2$ by the pair are given by 

\begin{equation}
    \partial_{c_1} \text{log(p}(a_{i'j'}=0)) =  \partial_{\alpha_{i'}} \text{log(p}(a_{i'j'}=0)),
\end{equation}
\begin{equation}
\partial_{c_1} \text{log(p}(a_{i'j'}=1)) =
    \frac{e^{C_1}(1+e^{C_1})^{-2}}{(1+e^{C_1})^{-1} - (1+e^{C_2})^{-1}},
\end{equation}
\begin{equation}
     \partial_{c_1} \text{log(p}(a_{i'j'}=2)) = 0.
\end{equation}
(For the second part, switch $i'$ and $j'$ for $\alpha$ and $\beta$ and consider $a_{j'i'}$.)
\begin{equation}
     \partial_{c_2} \text{log(p}(a_{i'j'}=0)) = 0,
\end{equation}

\begin{equation}
\partial_{c_2} \text{log(p}(a_{i'j'}=1)) = 
    -\frac{e^{C_2}(1+e^{C_2})^{-2}}{(1+e^{C_1})^{-1} - (1+e^{C_2})^{-1}},
\end{equation}

\begin{equation}
    \partial_{c_2} \text{log(p}(a_{i'j'}=2)) =  \partial_{\alpha_{i'}} \text{log(p}(a_{i'j'}=2)).
\end{equation}
(For the second part, switch $i'$ and $j'$ for $\alpha$ and $\beta$ and consider $a_{j'i'}$.)

\section{Bayesian correction of force term \label{ap:bayesian}}

For the inference of the model parameters we have only one sample. Although our model is a probabilistic model, our empirical distribution contains either $p(a_{ij}=1)=1$ if there is an edge, or $p(a_{ij}=1)=0$, if not. This can lead to divergences of the model parameters in the maximum likelihood solution. If one node is connected to all the other nodes, for instance, i.e. the graph contains a (N-1)-star subgraph, $\alpha_i$ will diverge in the maximum likelihood solution  ($\alpha_i \to \infty$). One way to avoid this problem is to formulate is as a Bayesian inference problem. This is a well-defined problem, even with a single data point. Lets denote the entries of the adjacency matrix of our graph $G$ as $a_{ij}$ and the corresponding random variables $A_{ij}$ or $\mathbf{A}$, respectively. Moreover, lets call the parameter vector of our model $\mathbf{\theta}=\{\mathbf{\alpha,\beta,x}\}$, with $\theta_i = \{ \alpha_i,\beta_i,\mathbf{x}_i \}$ and the corresponding random variable $X$. Then the posterior distribution of the parameters is given as
\begin{equation}
    p(X|\mathbf{A})=\frac{p(\mathbf{A}|X) p(X)}{p(\mathbf{A})}
    \label{eq:BayesianInference}
\end{equation}
Here $p(\mathbf{A}|X)$ is the likelihood (see Eq. (\ref{eq:likelihood_unweighted})), $p(X)$ comprises all prior distributions and $p(\mathbf{A})$ is the marginal likelihood of the data. Instead of asking for the parameters that maximise the likelihood we can now ask for the parameters that maximise the posterior $p(X|\mathbf{A})$ or, equivalently, its logarithm. If one considers the gradients of the logarithm of the posterior again as forces, the likelihood term produces the same forces as in the maximum likelihood case. However, we may get additional forces from the prior term, which, for instance, can prevent the activity parameters from diverging.\footnote{\cite{Barbera2015a} assumes normal priors for all parameters of the model. Nevertheless, it is noted in the main text that flat priors are used except for $\alpha$ and the positions $\mathbf{x}$. The mean of the prior distribution of $\alpha$ is set to $0$ and the prior distribution for the positions is $\mathcal{N}(0,1)$.}   

\section{Z-scores for Gaussian distribution of nodes }

\label{ap:zscore}

Table \ref{tab:zscore} gives the average z-scores for a Mantel test for the layout of a network generated from the Gaussian distribution of Fig. \ref{fig:validationsbm} -- two groups of nodes with a sigma of 1/12 and a distance of 5/6 between the groups with varying node number, averaged over three runs.
\begin{table}

\centering

\caption[Z-scores for Gaussian distributed nodes]{Average z-scores for a Mantel test for layout of a network generated from a Gaussian distribution of two groups of nodes with a sigma of 1/12 and a distance of 5/6 between the groups with varying node number (averaged over three runs).}
\label{tab:zscore}
\vspace{0.3cm}

\begin{tabular}{c|cccc } 
 \hline
 Node number & 100 & 300 & 600 & 900 \\ 
  \hline
 Unweighted & 36.65 & 77.49 & 90.56 & 94.91 \\ 
 Cumulative & 43.07 & 77.04 & 84.58 & 92.12\\
 Weighted & 45.60 & 83.32 & 92.84 & 95.69\\
 \hline
 \end{tabular}
\end{table}

\section{Real-world networks and comparison to other layout algorithms\label{ap:networks}}

\begin{figure}

    \includegraphics[width = \textwidth]{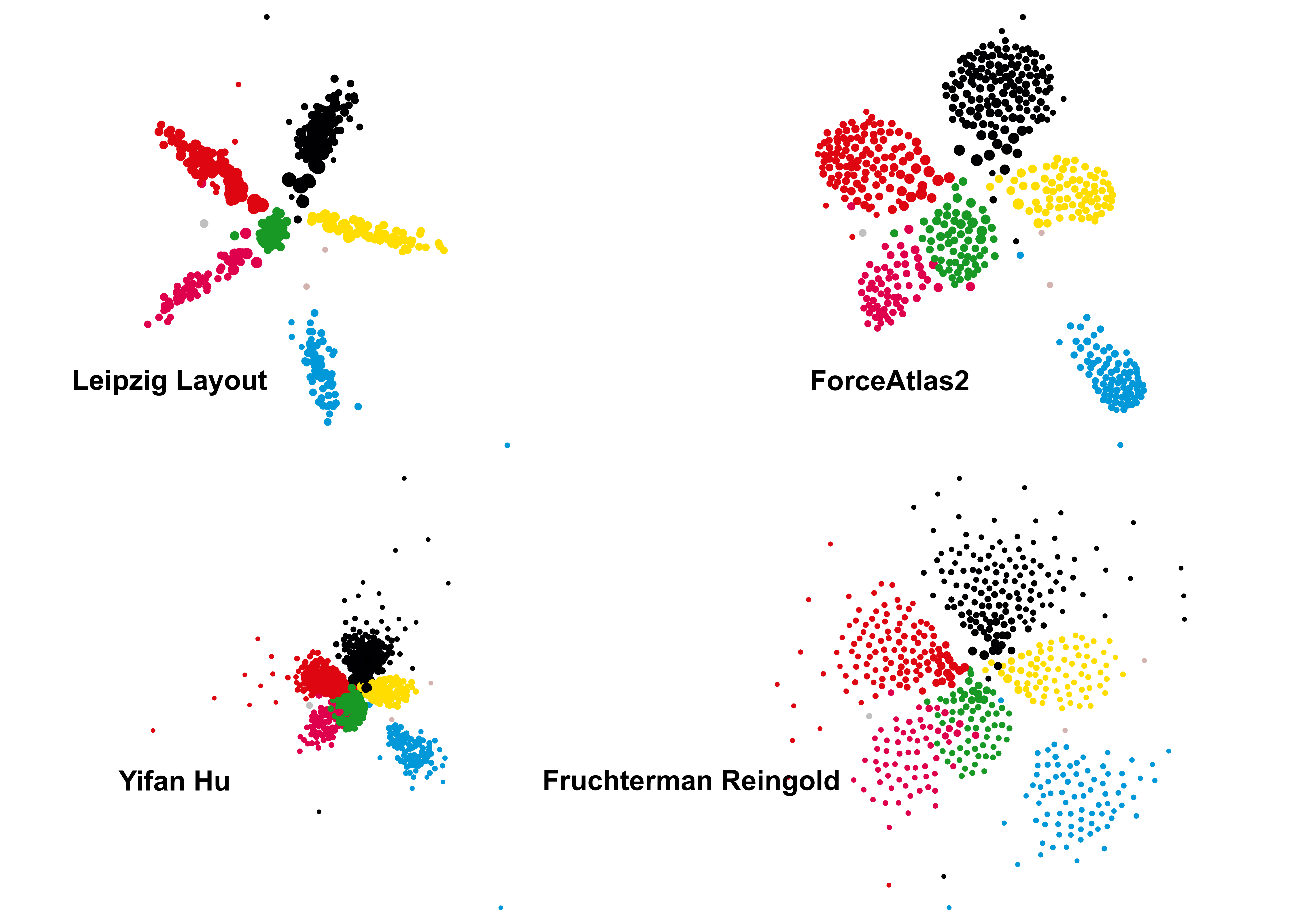}
    \caption[German parliament follower network in comparison with 3 other FDLs]{Bundestag follower network, comparison of Leipzig Layout (top left), ForceAtlas2 (top right), Yifan Hu (lower left), and Fruchterman Reingold (lower right). Overall placement of parties is similar, but Leipzig layout, which allows closer placement of nodes, arranges all parties (except the \textit{Greens}) along a one-dimensional axis.}
    \label{fig:bt-comp}
\end{figure}

\paragraph{German parliament}
In the main text, the Bundestag follower network layout was only compared to ForceAtlas2. In Fig. \ref{fig:bt-comp}, we also show the follower network spatialised with Yifan Hu \cite{hu2005efficient} (lower left) and Fruchterman Reingold \cite{fruchterman1991graph} (lower right). All layout algorithms roughly reproduce party divisions. Fruchterman Reingold tends to distribute nodes homogeneously in spaces. \textit{Die Linke} and the \textit{Greens} have some overlap in this layout. A less pronounced overlap is also visible for Yifan Hu, which produces a comparably dense spatialisation. 

\begin{figure}
   \noindent
  \makebox[\textwidth]{
    \includegraphics[width = 1.2\textwidth]{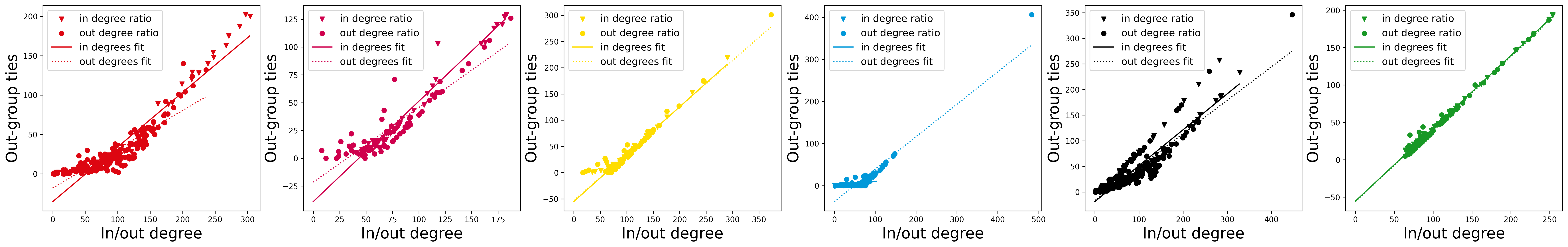}}
    \caption[Party in/out degree versus out-group ties]{Scatter plot of in/out degree and number of out-group ties for each party. No user from the \textit{Green} party has in in or out degree of less than 50. They use Twitter quite homogeneously, which explains their central placement in the force-directed layout (as well as the homogeneously distributed incoming and outgoing links among parties, see next Fig.).}
    \label{fig:bt-degrees}
\end{figure}

\begin{figure}

    \includegraphics[width = \textwidth]{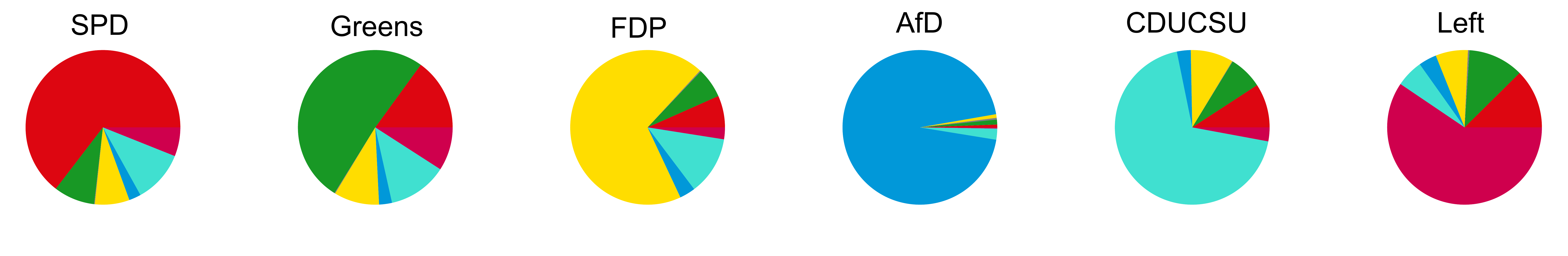}
    
    \vspace{1cm}
    \includegraphics[width = \textwidth]{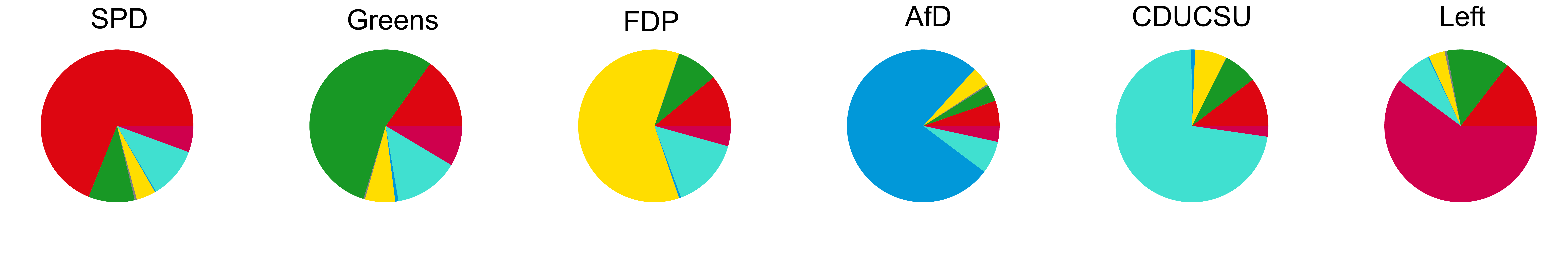}
    \caption[Incoming and outgoing links by party]{Incoming and outgoing links from/to different parties by party. \textit{Greens} are followed and follow all other parties (except \textit{AfD}) quite uniformly, which explains their central placement.}
    \label{fig:bt-inout}
\end{figure}

\begin{figure}
   \noindent
  \makebox[\textwidth]{
    \includegraphics[width = 1.2\textwidth]{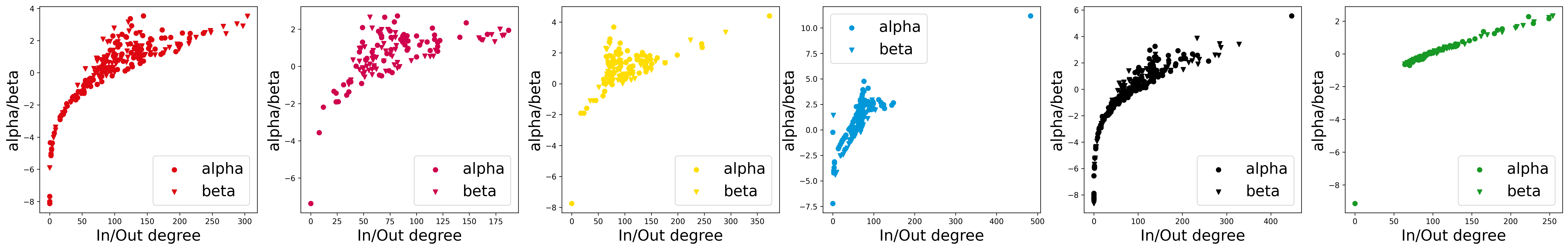}}
    \caption{Scatter plot of in/out degree and $\alpha$ and $\beta$ for each party.}
    \label{fig:bt-degrees-alphas}
\end{figure}

The central placement of the \textit{Greens} in the Leipzig Layout (as well as in the other layout algorithms) can be explained by their homogeneous usage of the platform: Each user of the party has an in/out degree of at least 50 (which is not the case for the other parties, see Fig. \ref{fig:bt-degrees}). Moreover, the party members are followed and follow all other parties in a quite uniform fashion (see Fig. \ref{fig:bt-inout}). 

Moreover, $\alpha$ and $\beta$ are correlated with node out and in degree for each party, see Fig. \ref{fig:bt-degrees-alphas}. In the case of the Greens, the correlation appears to be (except for one outlier) strongly linear.

\begin{figure}

    \includegraphics[width = 1\textwidth]{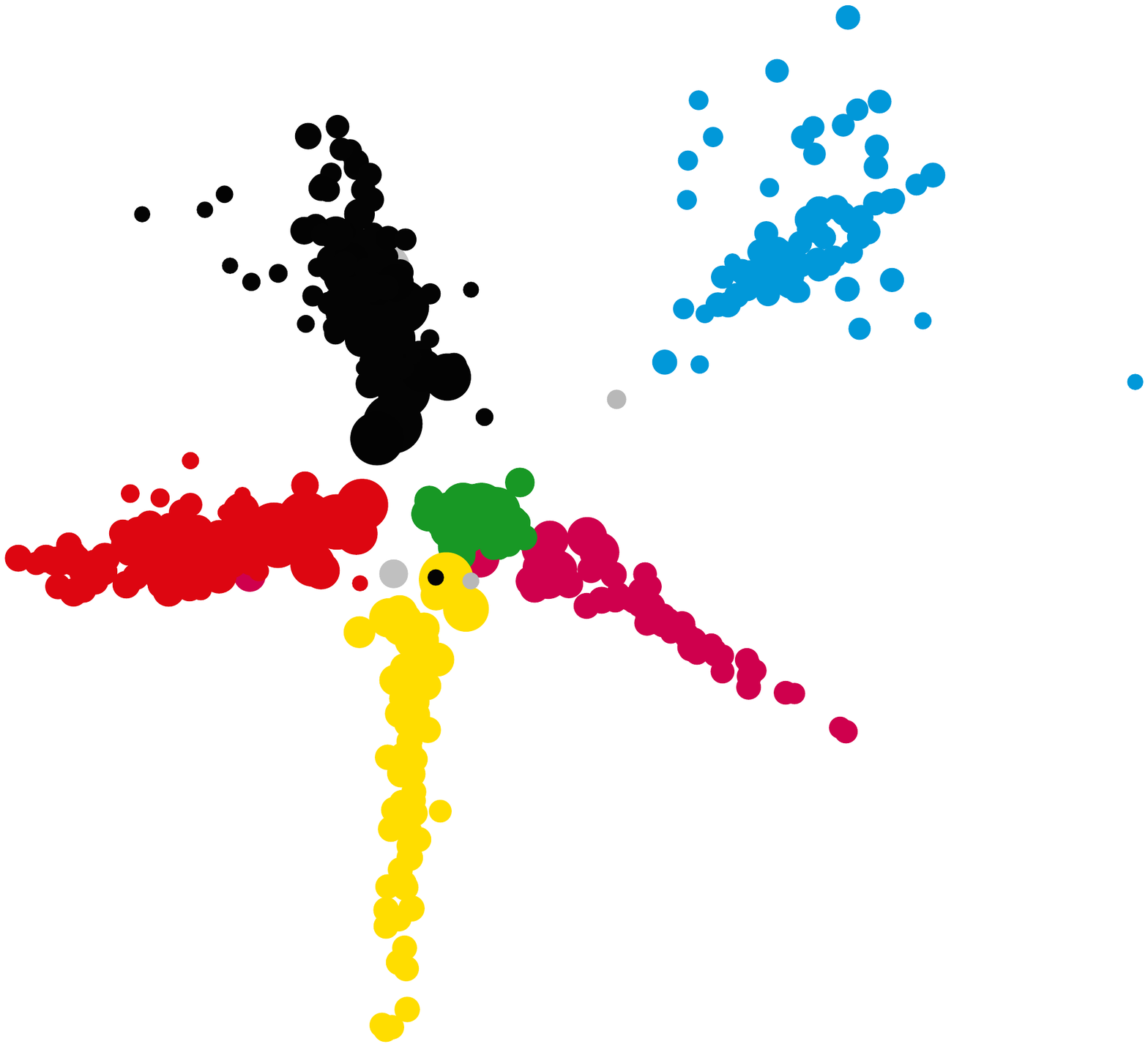}
    \caption[Local minimum of follower network]{Local minimum of the follower network of German deputies. Parties are still visibly separated, but \textit{FDP} is now placed between \textit{SPD} and \textit{Die Linke}. }
    \label{fig:bt-locmin}
\end{figure}

Different (local) minima exist for this network, one of which is displayed in Fig. \ref{fig:bt-locmin}. There, the \textit{FDP} is placed between \textit{SPD} and \textit{Die Linke}. The inference of local minima is a general problem of FDLs -- nevertheless, in the present framework, one can compare the log-likelihood of different equilibria and take the spatialisation with the highest likelihood (i.e. the lowest negative log-likelihood). The log-likelihood of Fig. \ref{fig:bt-locmin} is around 61,700, while it is roughly 57,700 for Fig. \ref{fig:bt-follow}. The minimum with the higher likelihood is also the politically more plausible one: \textit{SPD} and \textit{Die Linke} are, especially when it comes to economic policy, oftentimes strongly opposed to the market-liberal \textit{FDP}. On the other hand, \textit{FDP} and \textit{CDU/CSU} have often stressed that they are parties that have many things in common, such that Fig. \ref{fig:bt-follow} seems to be closer to political reality.

\paragraph{Harper's letter}

\begin{figure}

    \includegraphics[width = \textwidth]{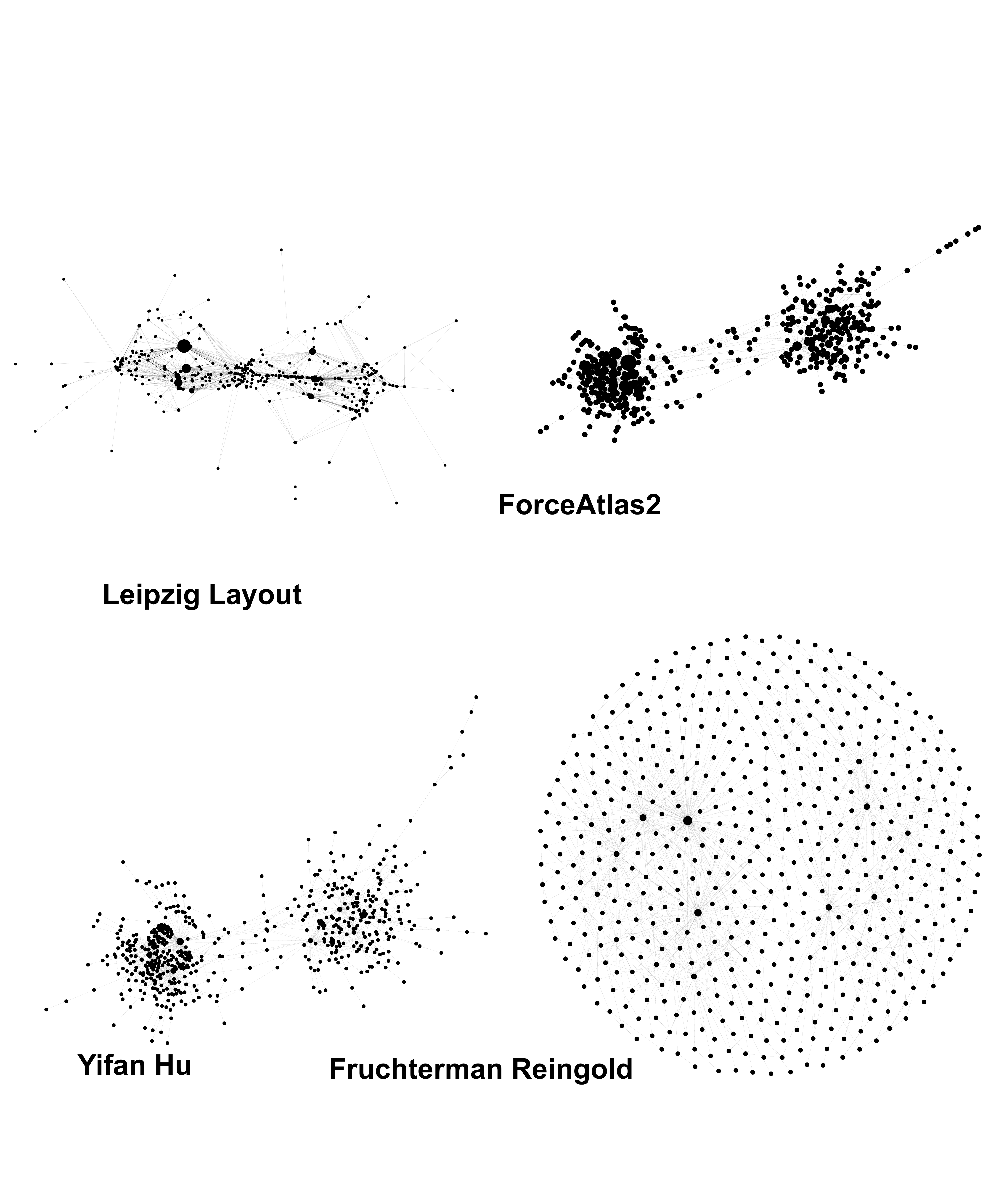}
    \caption[Harper's letter retweet network compared to three other FDLs]{Harper's letter retweet network comparison of Leipzig Layout (cumulative case, top left), ForceAtlas2 (top right), Yifan Hu (lower left), and Fruchterman Reingold (lower right).}
    \label{fig:harpers-comp}
\end{figure}

Fig. \ref{fig:harpers-comp} shows the Harper's letter retweet network spatialised with the four layout algorithms. Fruchterman Reingold, again, arranges the nodes rather uniformly in space. Interestingly, Yifan Hu and ForceAtlas2 produce a much more polarized spatialisation than Leipzig Layout.

\section{Survey on energy-generating technologies: Correlations \label{ap:correlations}}
\begin{figure}
\centering
\includegraphics[width = 0.8\textwidth]{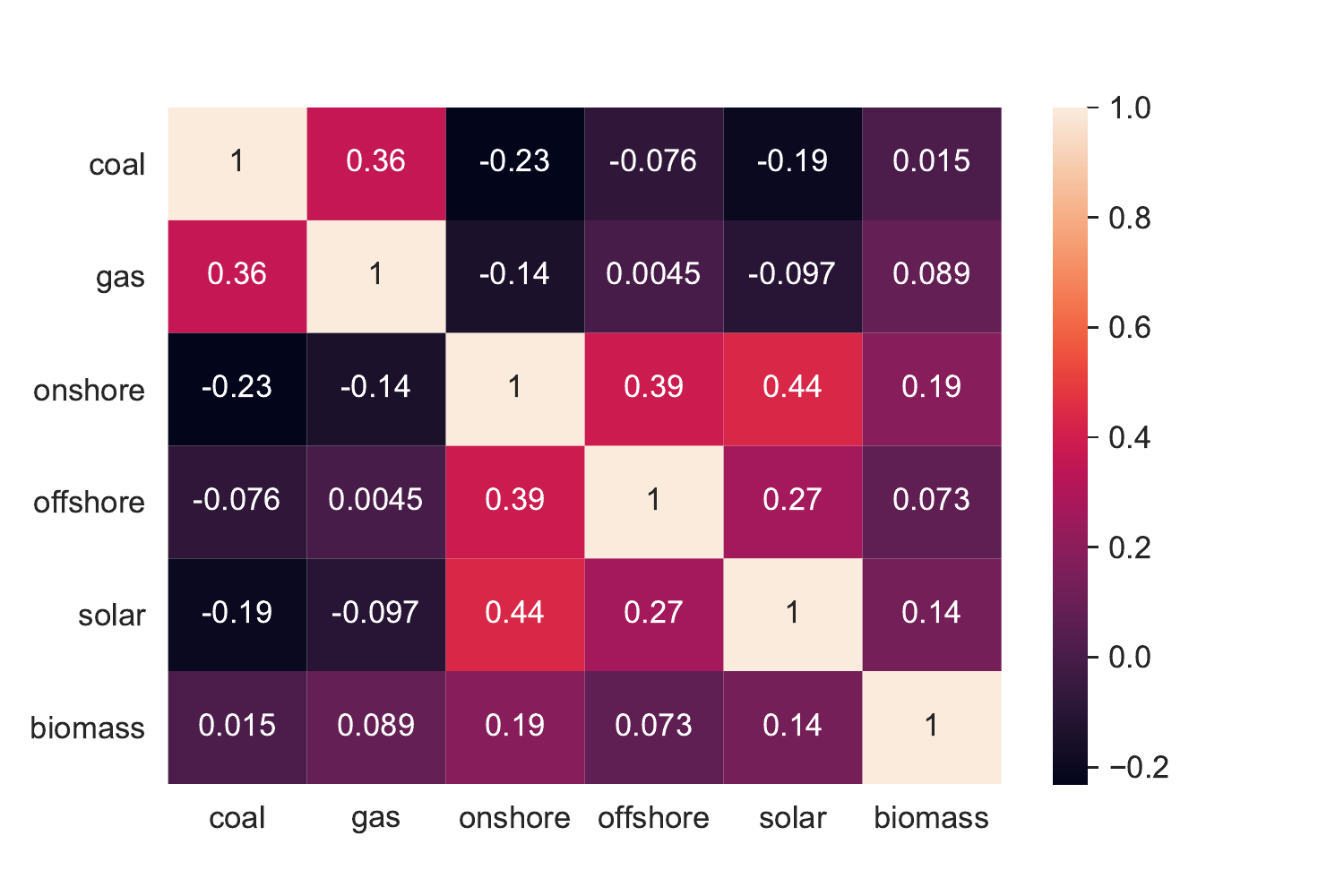}
    \caption[Survey correlations]{Correlations between ratings of different energy-generating technologies for the aggregated 3-point scale.}
    \label{fig:correlations}
\end{figure}

Pairwise Pearson correlation coefficients between ratings of different energy-generating technologies (for the aggregated 3-point scale) can be found in Fig. \ref{fig:correlations}. Solar and onshore technologies exhibit the strongest correlation. Gas and coal, as well as offshore and onshore technologies are also correlated relatively strongly. Biomass has, among the renewables, the weakest correlations with the other renewables. On the other hand, ratings are not negatively correlated with coal or gas. This is mirrored by its placement in Fig. \ref{fig:energy} at a certain distance from solar and onshore, and also offshore.

\end{document}